\documentclass{article} 
\usepackage{graphicx}
\usepackage{amsmath, amssymb,bbm}
\usepackage{hfstyle}

\newtheorem{proposition}{Proposition}
\newcommand{\ZZ}{{\mathbbm{Z}}} \newcommand{\NN}{{\mathbbm{N}}}
 \newcommand{\F}{{\cal F}}
\newcommand{\B}{{\cal B}} \newcommand{\G}{{\cal G}}
\newcommand{\calS}{{\cal S}} \newcommand{\f}{{\rm \bf f}}

\newcommand{\card}{\mathop{\rm{card}}}

\begin{document}
\title{Conservation and dynamics of blocks 10 in the cellular automaton rule 142}
\author{Henryk Fuk\'s
      \oneaddress{
         Department of Mathematics, Brock University\\
         St. Catharines, Ontario  L2S 3A1, Canada\\
         Toronto, Ontario M5T 3J1, Canada\\
         \email{hfuks@brocku.ca}
       }
   }

\Abstract{
We investigate dynamics of the cellular automaton rule 142.
This rule possesses additive invariant of the second order, namely
it conserves the number of blocks $10$. Rule 142 can be alternatively
described as an operation on a binary string in which we
simultaneously flip all symbols which have dissenting right
neighbours.
We show that the probability of having a dissenting neighbour
can be computed exactly using the fact that the surjective rule
60 transforms rule 142 into rule 226. We also demonstrate that the  
conservation of the number of $10$ blocks implies that these
blocks move with speed $-1$ or stay in the same place, depending on
the state of the preceding site. At the density of blocks
$10$ equal to 0.25, the rule 142 exhibits a phenomenon
similar to the jamming transitions occurring in discrete models
of traffic flow.
}

%%% ----------------------------------------------------------------------
\maketitle

\section{Introduction}
Let $s$ be a binary string of $L$ symbols, i.e., $s=s_0s_1\ldots
s_{L-1}$, where $s_i \in \{0,1\}$ for $0 \leq i \le L$, $L\in \NN$. We will say
that the symbol $s_i$ has a \emph{dissenting right neighbour} if
$s_{i-1}= s_i\neq s_{i+1}$. By \emph{flipping} a given symbol $s_i$ we
will mean replacing it by $1-s_i$.

Consider now the following problem: suppose that we
simultaneously flip all symbols which have dissenting right
neighbours, as shown in the example below. 
\begin{equation} \label{flipp}
\begin{array}{cccccccccccccc}
   \cdots&0&0&1&0&0&1&1&1&0&1&0&1&\cdots\\
    & &\Big\downarrow& & &
        \Big\downarrow & & &
             \Big\downarrow& & & & & \\  
   \cdots&0&1&1&0&1&1&1&0&0&1&0&1&\cdots
\end{array} 
\end{equation} 
Assuming that the initial string is randomly generated, what is 
the probability that a given symbol has a dissenting right neighbour
after $t$ iterations of the aforementioned procedure?

In order to answer this question, we will take advantage of the fact
that the process described in the previous paragraph is actually a
cellular automaton rule 142, using Wolfram's numbering scheme for
elementary cellular automata \cite{wolfram2002}. It has a property which turns out to be
crucial to the solution. If one counts the number of pairs $10$ in
(\ref{flipp}) before and after the bit-flipping operation, it is easy
to see that this number remains constant (similarly as the number of pairs
$01$). We will show that this is
true for an arbitrary string, and we will take advantage of this fact to
compute the probability of having a dissenting neighbour.

\section{Definitions}
Before we proceed, we will introduce several concepts of cellular
automata theory.
Let $\G=\{0,1\}$ be called {\it a symbol set}, and
$\calS=\{0,1\}^{\ZZ}$ be called {\it the configuration space}.  {\it A
  block of radius} $r$ is an ordered set $b_{-r} b_{-r+1} \ldots b_r$,
where $r\in \NN$, $b_i \in \G$. Let $r\in \NN$ and let $\B_r$ denote
the set of all blocks of radius $r$ over $\G$. The number of elements
of $\B_r$ (denoted by $\card \B_r$) equals $2^{2r+1}$.

A mapping $f:\{0,1\}^{2r+1}\mapsto\{0,1\}$ will be called {\it a
  cellular automaton rule of radius $r$}. Alternatively, the function
$f$ can be considered  a mapping of $\B_r$ into $\B_0=\G=\{0,1\}$.

Corresponding to $f$ (also called {\it a local mapping}), we define a
{\it global mapping} $F:S\to S$ such that $
(F(s))_i=f(s_{i-r},\ldots,s_i,\ldots,s_{i+r}) $ for any $s\in S$. The
{\it composition of two rules} $f,g\in \F$ can be now defined in terms
of their corresponding global mappings $F$ and $G$ as $(F\circ
G)(s)=F(G(s)),$ where $s \in S$. We note that if $f \in \F_p$ and $g
\in \F_q$, then $f \circ g \in \F_{p+q}$. For example, the composition
of two radius-1 mappings is a radius-2 mapping:
\begin{equation}
  (f\circ g)(s_{-2},s_{-1},s_0,s_1,s_2)=
  f(g(s_{-2},s_{-1},s_0),g(s_{-1},s_0,s_1),g(s_0,s_1,s_2)).
\end{equation} Multiple composition will be denoted by
\begin{equation}f^n=\underbrace{f \circ f \circ \cdots \circ
    f}_{\mbox{$n$ times}}.
\end{equation}

A {\it block evolution operator} corresponding to $f$ is a mapping
$\f:\B \mapsto \B$ defined as follows. Let $r \geq p >0$, $a\in \B_r$,
$f\in \F_p$, and let $b_i=f(a_{i-p},a_{i-p+1}, \ldots,a_{i+p})$ for
$-r+p \leq i \leq r-p$. Then we define $\f(a)=b$, where $b \in
\B_{r-p}$. Note that if $b \in B_1$ then $f(b)=\f(b)$.

In this paper, we will be concerned with trajectories of a given
configuration under consecutive iterations of $F$. Denoting the
initial configuration by $s(0)$, the image of $s(0)$ after $t$
iterations of $F$ will be denoted by $s(t)$, i.e.,
\begin{equation}
  s(t)=F^t(s(0)),
\end{equation}  
which imples that
\begin{equation}
  s(t+1)=F(s(t)),
\end{equation} 
and hence
\begin{equation} \label{cadef} s_i ( t + 1 ) = f ( s_{i - r} ( t ),
  s_{i -r + 1} ( t ), \ldots, s_{i + r} ( t ) ) .
\end{equation}

Cellular automaton rule 142, which is the subject of this paper, has
the following local function
\begin{eqnarray}\label{rletable}
  f(0,0,0)=0,
  f(0,0,1)=1,
  f(0,1,0)=1,
  f(0,1,1)=1,\\ \nonumber
  f(1,0,0)=0,
  f(1,0,1)=0,
  f(1,1,0)=0,
  f(1,1,1)=1,
\end{eqnarray}
which can also be written in an algebraic form 
\begin{equation} \label{r142} f(x_0,x_1,x_2)=x_1+
(1-x_0)(1-x_1)x_2 - x_0x_1(1-x_2).
\end{equation}  

\section{Conservation}
As shown in \cite{paper23}, rule 142 is one of the few nontrivial
elementary rules which posses the second order additive invariant. It
conserves the number of blocks $10$, and this fact can be formally
described as follows. Let us first define a function
$\xi(x_0,x_1)=x_0(1-x_1)$, which takes value $1$ on block $10$ and
value $0$ on all other blocks of length $2$. We will call $\xi$ the
\emph{density} of blocks $10$. Following \cite{Hattori91}, we will say that $\xi$ is a density
function of an additive invariant of $f$ if
\begin{equation} \label{conscond} \sum_{i=0}^{L-1} \xi(f( s_i,
  s_{i+1}, s_{i+2}), f(s_{i+1}, s_{i+2},s_{i+3}))= \sum_{i=0}^{L-1}
  \xi(s_i,s_{i+1})
\end{equation} 
for every positive integer $L$ and for all $s_0,s_1,\ldots, s_{L-1}
\in \{0,1\}$. In the above, and in all subsequent considerations, we
are assuming that addition of all spatial indices is performed
\emph{modulo} $L$. That is, we will be concerned with periodic
configurations, or, in other words, configurations with periodic
boundary conditions where $s(i+L)=s(i)$ for all integers $i$.

The right hand side of the above equation simply denotes the number of
blocks $10$ in the configuration $s=(s_0,s_1,\ldots, s_{L-1})$, and the
left hand side denotes the number of these blocks in the image of $s$
under $f$.
\begin{figure}
  \begin{center}
    \includegraphics[scale=1.0]{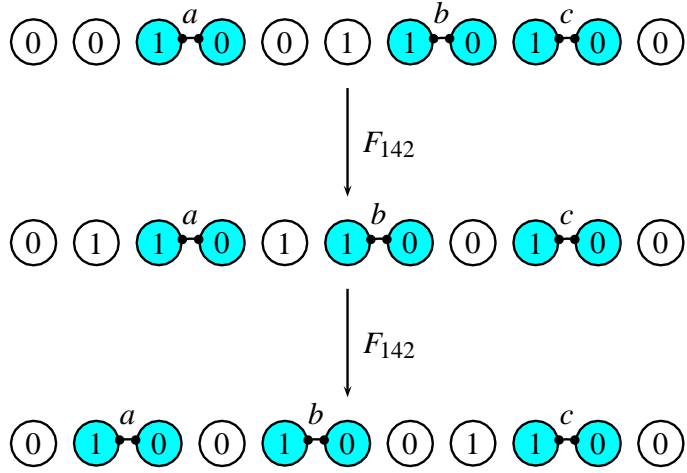}
  \end{center}
  \caption{Two consecutive images of a sample configuration under rule
    142, assuming periodic boundary conditions. Blocks $10$ are
    labelled with letters a,b,and c.}
  \label{fig1}
\end{figure}
Figure~\ref{fig1} shows an example of a configuration $s$ consisting
of 11 sites, and its two consecutive images under rule 142, i.e.,
$F_{142}(s)$ and $F^2_{142}(s)$, where $F_{142}$ denotes the global
function of rule 142. Periodic boundary conditions are assumed. The
initial configuration $s$ contains three blocks $10$ labelled $a$,
$b$, and $c$, and one can clearly see that the number of blocks $10$
remains constant after each application of $F_{142}$.  Moreover, since
the number of blocks $10$ remains constant, we can label them with
distinctive labels, which will allow us to keep track of individual
blocks. In  Figure~\ref{fig1}, such labelling can be simply  obtained 
by enumerating blocks $10$ from left to right\footnote{In general, such enumeration will not be unique, because
we have to decide which block is first. However, by imposing some additional
conditions, it is possible to obtain unique labelling. An algorithm producing
such labelling has been described in \cite{paper10}}.
For example, looking again at Figure~\ref{fig1}, we could say
that block $a$ remains in the same position after the first iteration,
but moves to the left by one site in the second iteration.  Similarly,
block $b$ moves by one site to the left in both iterations shown in
Figure~\ref{fig1}.

To formalize the concept of the motion of blocks, we will first prove
that rule 142 conserves the number of blocks $10$ in an arbitrary
periodic configuration. Let us note that
\begin{align}
  \xi(f( s_i, s_{i+1}, s_{i+2}), f(s_{i+1}, s_{i+2},s_{i+3}))=\\ \nonumber
  f( s_i, s_{i+1}, s_{i+2}) (1-f(s_{i+1}, s_{i+2},s_{i+3})).
\end{align} 
Using (\ref{r142}), the right hand side of the above equation becomes
somewhat complicated, but it drastically simplifies if one notes that
all variables $s_i$ in this equation are Boolean, and $x^n=x$ for all
positive $n$ if $x\in\{0,1\}$. After this simplification, one obtains
\begin{align}
  \xi(f( s_i, s_{i+1}, s_{i+2}), f(s_{i+1},& s_{i+2},s_{i+3}))= \nonumber\\
  - s_i s_{i+1} + s_{i+1}+&  s_i s_{i+1} s_{i+2}  - s_{i+1} s_{i+2} s_{i+3}.
\end{align} 
We will now regroup terms on the right hand side
\begin{align}
  \xi(f( s_i, s_{i+1}, s_{i+2}), f(s_{i+1}, s_{i+2},s_{i+3}))&=\nonumber\\
  s_{i+1} (1-s_{i+2}) + s_{i+1} s_{i+2} (1-s_{i+3}) 
  &-s_{i} s_{i+1} (1-s_{i+2}),
\end{align} 
and finally write the last equation as
\begin{align} \label{condiscr} \xi(f( s_i, s_{i+1},& s_{i+2}),
  f(s_{i+1}, s_{i+2},s_{i+3}))=\\
   &\xi(s_{i+1} ,s_{i+2})) - J(s_{i+1},
  s_{i+2}, s_{i+3}) + J( s_{i}, s_{i+1} ,s_{i+2}),\nonumber
\end{align} 
where
\begin{equation} \label{currentdefinition} J(x_0,x_1,x_2)=x_0 x_1
  (x_2-1).
\end{equation} 
This leads to
\begin{multline}
  \sum_{i=0}^{L-1} \xi(f( s_i, s_{i+1}, s_{i+2}), f(s_{i+1},
  s_{i+2},s_{i+3}))=
  \sum_{i=0}^{L-1} \xi(s_{i+1} ,s_{i+2}) \\
  - \sum_{i=0}^{L-1} J(s_{i+1}, s_{i+2}, s_{i+3}) + \sum_{i=0}^{L-1}
  J( s_{i}, s_{i+1} ,s_{i+2}),
\end{multline}
and, since $\sum_{i=0}^{L-1} J(s_{i+1}, s_{i+2}, s_{i+3}) =
\sum_{i=0}^{L-1} J( s_{i}, s_{i+1} ,s_{i+2})$, the conservation
condition (\ref{conscond}) follows.

The equation(\ref{condiscr}) resembles continuity equation, with $J$
playing a role of current, or flow of blocks $10$ \cite{Hattori91}. To see this, let
us note that $J(s_i(t),s_{i+1}(t), s_{i+2}(t))$ takes non-zero value
only when $s_i(t)$,$s_{i+1}(t)$, $s_{i+2}(t)=1,1,0$. Consider now a
configuration containing block $110$, surrounded by sites of
undetermined state.  Using definition of the local function for rule
142 (\ref{rletable}), we can construct partial state of the
configuration at the next time step. Denoting by $*$ an arbitrary
value in the set $\{0,1\}$, we have $f(*,1,1)=1$, $f(1,1,0)=0$, and
$f(1,0,*)=0$, hence
\begin{align}
  s(t)=&\ldots\mathtt{110}\ldots\\ \nonumber
  s(t+1)=&\ldots\mathtt{100}\ldots
\end{align}
We can clearly see that the block $10$, when preceded by 1,
moves by one site to the left in a single iteration. Similar argument
could be used to demonstrate that the block $10$ preceded by 0
does not move:
\begin{align}
  s(t)=&\ldots\mathtt{010}\ldots\\ \nonumber
  s(t+1)=&\ldots\mathtt{?10}\ldots,
\end{align}
where ``$?$'' denotes undetermined value. Additionally,
note there is no other way to obtain the block $10$ in $s(t+1)$,
that is, if $s_i(t+1)s_{i+1}(t+1)=10$, then we must have 
either $s_{i}(t)s_{i+1}(t)s_{i+2}(t)=110$ or
$s_{i-1}(t)s_{i}(t)s_{i+1}(t)$ $=010$.
This demonstrates that indeed only
blocks $110$ can contribute to the current, in agreement with eq.
(\ref{currentdefinition}).
 
\section{Initial distribution}
Let us now go back to the problem stated in the introduction.  In
order to make the problem well posed, we need to define the
probability distribution $\mu$ from which the initial string is drawn.
Since we know that the rule 142 conserves the number of blocks $10$,
it is natural to consider an initial distribution parameterized by the
density of blocks $10$.  Let us define the expected value of $\xi$
at site $i$ as
 \begin{equation} \label{defdens} \rho(i,t)=E_\mu \left[
     \xi(s_i(t),s_{i+1}(t)) \right] = E_\mu \left[s_i(t)
     (1-s_{i+1}(t))\right].
 \end{equation}
 Assuming that the initial distribution $\mu$ is translation-invariant,
 $\rho(i,t)$ will not depend on $i$, and we will therefore define
 $\rho(t)=\rho(i,t)$.  Furthermore, since $\xi$ is density function of
 a conserved quantity, $\rho(t)$ is $t$-independent, so we define
 $\rho=\rho(t)$.

 The desired distribution parameterized by $\rho$ can be obtained as
 follows.  Let $\rho \in [0,1/2]$ be the target density of blocks
 $10$, and let $\{X_i\}_{i=0}^{L-1}$ be a collection of identical independently distributed
 Bernoulli random variables such that
\begin{align} \label{factor2}
  Pr(X_i=1)&=2 \rho,\\
  Pr(X_i=0)&=1-2 \rho,
\end{align}
for all $i\in\{0,1\}$. The initial configuration will be given by
\begin{equation} \label{initdist} s_i(0)=\left( \sum_{j=0}^i
    X_j\right)\mod 2.
\end{equation}
Note that when $X_i == 1$, we obtain either a subsequence $s_i(0)s_{i+1}(0)=01$ or
$s_i(0)s_{i+1}(0)=10$. Those subsequences occur at the 
same frequency, which accounts for the factor of 2 in eq. (\ref{factor2}).

Let $P_t(b)$ denote the probability of occurrence of block the $b$ in
the configuration $s(t)$.  If the density of blocks $10$ in the
initial configuration is $\rho$, then the probability of having a
dissenting neighbour at time $t$ will be denoted by
$P_\mathrm{dis}(\rho,t)$.  A site $s_i$ has a dissenting right
neighbour if $s_{i-1}s_is_{i+1}=110$ or $s_{i-1}s_is_{i+1}=001$.
$P_\mathrm{dis}(\rho,t)$ is therefore given by
\begin{equation}
  P_\mathrm{dis}(\rho,t)=P_t(110)+P_t(001).
\end{equation}
Although two block probabilities appear on the right hand side of the
above definition, we will show that $P_\mathrm{dis}(\rho,t)$ can be
expressed in terms of a single block probability.

As a first step, we note that the following properties are direct
consequences of the definition (\ref{initdist}).
\begin{proposition} \label{propertiesofP} Let $P_o(b)$ denotes the
  probability of occurrence of block $b$ in the configuration drawn
  from the distribution given by (\ref{initdist}). Then we have:
  \begin{itemize}
  \item[(i)] $P_0(1)=P_0(1)=1/2$
  \item[(ii)] $P_0(10)=P(01)=\rho$
  \item[(iii)] $P_0(b)=P_0(\bar{b})$, where $\bar{b}$ denotes Boolean
    conjugation of block $b$, i.e. $\bar{b}_i=1-b_i$.
  \end{itemize}
\end{proposition}
Rule142 exhibits Boolean self-conjugacy, that is, replacing all zeros
by ones and vice versa in the definition (\ref{rletable}) does not
change the definition. This fact together with
Proposition~\ref{propertiesofP}(iii) implies that $P_t(110)=P_t(001)$,
hence
\begin{equation} \label{pdis1}
  P_\mathrm{dis}(\rho,t)=2P_t(110).
\end{equation}
 Kolmogorov consistency conditions for block probabilities require that
\begin{eqnarray*}
  P_t(110)+P_t(111)&=&P_t(11),\\
  P_t(10)+P_t(11)&=&P_t(1),
\end{eqnarray*}
hence
\begin{equation}
  P_t(110)=P_t(1)-P_t(10)-P_t(111).
\end{equation} 
Using the fact that $P_t(10)=\rho$, we obtain
\begin{equation} \label{Probdis1} 
 P_\mathrm{dis}(\rho,t)=2P_t(1)-2\rho-2P_t(111).
\end{equation}
The next result will lead to the elimination of $P_t(1)$ from
the above equation.
\begin{proposition}\label{onehalf}
  Let the initial configuration $s(0)$ be drawn from the distribution
  given by~(\ref{initdist}), and let $s(t)$ be obtained from $s(0)$ by
  iterating rule 142 $t$ times, so that $s(t)=F_{142}^t(s(0))$. Then we
  have
  \begin{equation}
    P_t(1)=P_t(0)=1/2.
  \end{equation}
\end{proposition}
We will prove this by induction. Obviously, $P_0(1)=P_0(0)=1/2$ by
Proposition~\ref{propertiesofP}.  Let us assume that $P_t(1)=1/2$ for
some $t$. Block $1$ has four preimages under $\mathbf{f}_{142}$, and
these are $001$, $010$, $011$, $111$. This leads to
\begin{equation}
  P_{t+1}(1)=P_t(001)+P_t(010)+P_t(011)+P_t(111).
\end{equation} 
Kolmogorov consistency conditions require that
$P_t(011)+P_t(111)=P_t(11)$, and, as remarked before, Boolean
self-conjugacy of the rule 142 implies $P_t(001)=P_t(110)$.
This yields
\begin{equation}
  P_{t+1}(1)=P_t(110)+P_t(010)+P_t(11).
\end{equation}  
Using consistency conditions again we get
\begin{equation}
  P_{t+1}(1)=P_t(10)+P_t(11)=P_t(1),
\end{equation}  
and this, by induction hypothesis, yields $P_{t+1}(1)=1/2$, concluding
the proof.

Proposition~\ref{onehalf} simplifies eq. (\ref{Probdis1}) to
\begin{equation}  \label{Probdis2}
 P_\mathrm{dis}(\rho,t)=1-2\rho-2P_t(111).
\end{equation}
Now the only thing left is to compute the probability of occurrence of
block 111 in the configuration~$s(t)$.

\section{Preimages}
In order to compute $P_t(111)$, we will use some properties of
preimages of the block $111$. Let $\mathbf{f}^{-1}_{142}(111)$ be a
set of preimages of $111$ under $\mathbf{f}_{142}$. Then we have
\begin{equation}
  P_t(111)=\sum_{b\in \mathbf{f}^{-1}_{142}(111)} P_{t-1}(b).
\end{equation} 
Generalizing the above, we can write
\begin{equation}
  P_t(111)=\sum_{b\in \mathbf{f}^{-t}_{142}(111) } P_{0}(b),
\end{equation} 
where again $\mathbf{f}^{-t}_{142}(111)$ is a set of preimages of
$111$ under $\mathbf{f}_{142}^t$, i.e., under $t$ iterations of
$\mathbf{f}_{142}$. To find $P_t(111)$ using the above property, two
steps are needed: first, we have to find the set of preimages of
$111$, and then to find probabilities of their occurrences in the
initial distribution.
\begin{figure}
   \begin{flushleft}
  \includegraphics[scale=0.6]{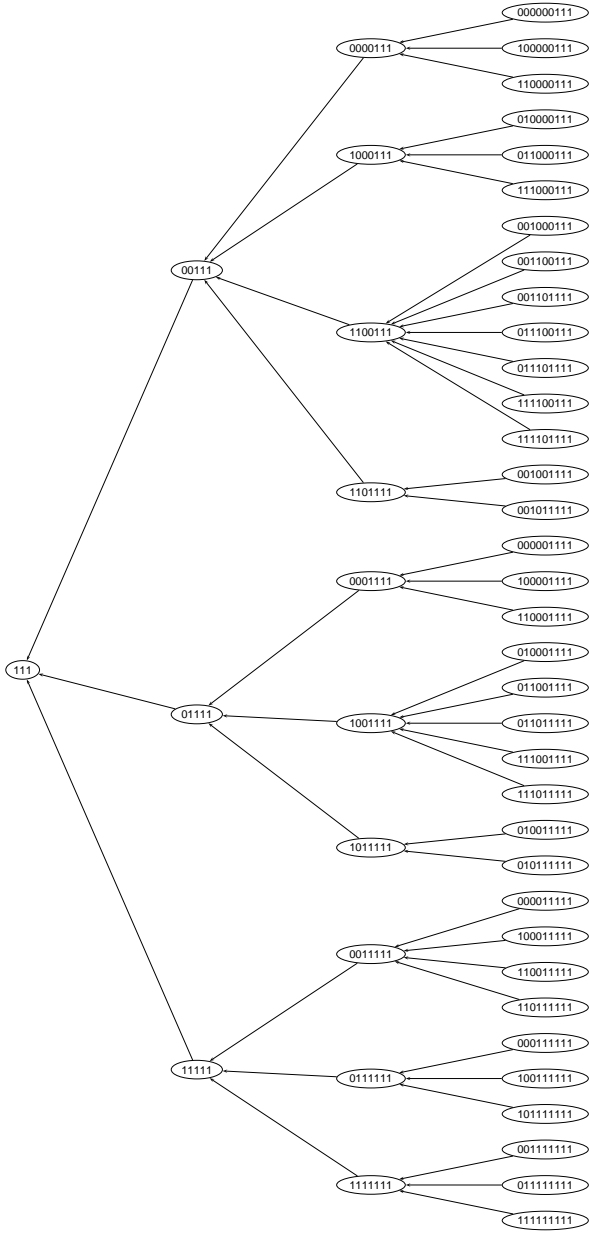}
   \end{flushleft}
  \caption{Tree of preimages of the block $111$ under rule 142.}
  \label{fig2}
\end{figure}
Figure~\ref{fig2} shows three levels of preimages of $111$. Upon
inspection of this figure, two properies become apparent.
\begin{proposition}
  Let $b$ be a $t$-step preimage of 111'', that is, $b \in
  \mathbf{f}_{142}^{-t}(111)$. Then
  \begin{itemize}
  \item[(i)] The length of $b$ is $3+2t$;
  \item[(ii)] $b$ ends with $111$.
  \end{itemize}
\end{proposition}

The first property is an obvious consequence of the definition of
$\mathbf{f}_{142}^t$, and the second one can be easily proved by
induction (omitted here).

Further inspection of Figure~\ref{fig2} leads to the necessary and
sufficient condition for a block $b$ to be a $t$-step preimage of
$111$. Before stating this condition formally, we will explain it
using an example. 
 
 Consider the block $b=011100111$, which is a
preimage of $111$ in three steps since
$\mathbf{f}_{142}(011100111)=1100111$,
$\mathbf{f}_{142}(1100111)=00111$, and $\mathbf{f}_{142}(00111)=111$.
Let us now assume that we start with a ``capital'' of $1$. We will move
along the string $b=b_0b_1\ldots b_8$ starting from $i=6$ and moving
in the direction of decreasing $i$.  Every time we see that $b_{i-1}$
is different from $b_i$, we decrease out ``capital'' by $1$. If
$b_{i-1}=b_i$, we increase our ``capital'' by~$1$.  We stop at $i=1$.

Clearly, it is possible to traverse $b=011100111$ following this
procedure without making the capital negative. It turns out that this
is a general property of preimages of $111$. If $b$ is a preimage of
$111$, then it is possible to traverse it keeping the capital
non-negative.  If $b$ is not a preimage of $111$, the capital will
become negative at some point. A more formal statement of this
property is as follows.
\begin{proposition} \label{preimenum} Let $t$ be a non-negative
  integer, and let $b=b_0b_1\ldots b_{2t+2}$ be a binary string of
  length $3+2t$ ending with $111$. Define $\chi$ to be a function of
  two variables such that $\chi(u,v)=1$ if $u=v$, and $\chi(u,v)=-1$
  otherwise.  The string $b$ is a preimage of $111$ under
  $\mathbf{f}_{142}^t$ if and only if the inequality
  \begin{equation} \label{preimcondition} 1+\sum_{i=0}^{k}
    \chi(b_{2t-i-1},b_{2t-i}) \geq 0
  \end{equation}
  is satisfied for all $k=0,1,\ldots,2t-1$.
\end{proposition}

Instead of proving this proposition directly, we will show that it can
be derived from a similar result previously obtained for a related
cellular automaton rule.
\section{Rule 226}
\begin{figure}
  \begin{center}
\includegraphics[scale=1]{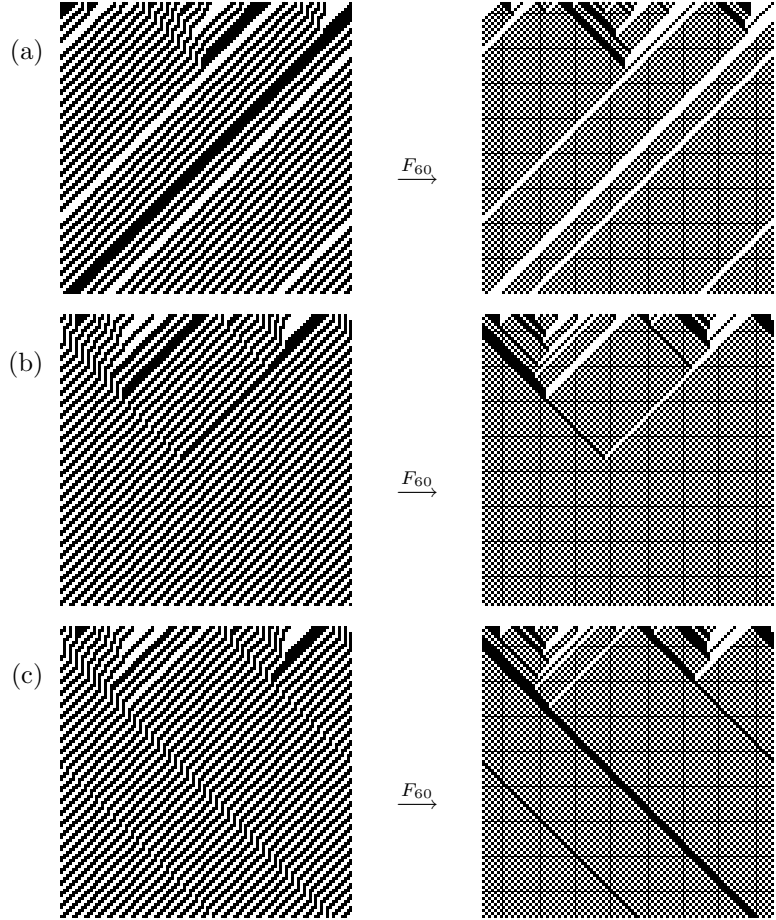} 
  \end{center}
  \caption{Spatiotemporal patterns generated by the rule 142 and their transformations under the rule 60. Densities of 
  blocks $10$ are (a) $\rho=0.23$, (b) $\rho=0.25$, and (c) $\rho=0.26$. All patterns show 100 consecutive
  iterations of a lattice of 100 sites with periodic boundary conditions. Black squares represent $1$, white
  spaces represent $0$.}
  \label{fig3}
\end{figure}

In \cite{Boccara93} it has been observed that
\begin{equation} \label{r142trafo} f_{60} \circ f_{142} = f_{226}
  \circ f_{60},
\end{equation}
where
\begin{align}
  f_{60}(x_0,x_1,x_2)&=x_0+x_1 \mod 2\\
  f_{226}(x_0,x_1,x_2)&=x_0 x_1 -x_1 x_2+ x_2.
\end{align}
This means that there exists a local mapping (rule 60) which
transforms rule 142 into rule 226. The above correspondence 
between rules 142 and 226 has been illustrated in Figure~\ref{fig3}, which
shows spatiotemporal patters of rule 142 (left panel) as well as images of these patterns
under the rule 60 (right panel). The patterns in the right panel are in fact identical to spatiotemporal patterns
which one would obtain by iterating rule 226, providing that the
initial configuration in the right panel has been obtained by applying rule 60
to the corresponding initial configuration from the left panel. 

 Rule 226 and its image under
spatial reflection, rule 184, are the only non-trivial elementary
number-conserving rules, and many results regarding their dynamics
have been established \cite{Krug88,Nagatani95,Nagel96,Belitsky98,Sipper96,paper11,NishinariT98,wolfram2002,Blank03}.
For our purpose, one such result will be
particularly useful.
\begin{proposition} \label{fuks97prop} Under the rule 226, $t$ step
  preimages of 00 have the following properties:
  \begin{itemize}
  \item[(i)] In each preimage, the number of zeros exceeds the number
    of ones.
  \item [(ii)] The block $a_0 a_1 \ldots a_{2t+1}$ is an $t$-step
    preimage of $00$ if and only if it ends with two zeros and
    \begin{equation}
      1+\sum_{i=2}^{k} \xi(s_{2t+1-i}) \geq 0
    \end{equation}  for every
    $2 \leq k \leq 2t+1$, where $\xi(0)=1$, $\xi(1)=-1$.
  \item[(iii)] The number of $t$-step preimages of 00 containing
    exactly $n_0$ zeros and $n_1$ ones is equal to
    \begin{equation}
      \frac{n_0-n_1}{n_0+n_1}  \binom{n_0+n_1}{n_1},
    \end{equation}
    where $n_0+n_1=2t+2$.
  \end{itemize}
\end{proposition}
Proof of this result can be found in \cite{paper4}, with further
generalization in \cite{paper11}. The proof is based on the fact that
the enumeration of preimages of $00$ under rule 226 (or 184) is
equivalent to the problem of enumeration of planar lattice paths
between two points, subject to some constraining conditions. This path
enumeration problem can then be solved using combinatorial methods.

We will now relate preimages of rules 226 and 142.
\begin{proposition} \label{equival} The number of $t$ step preimages
  of {\tt 111} under rule 142 is equal to the number of $t$ step
  preimages of {\tt 00} under rule 226.
\end{proposition}

We will explicitly construct a bijection $T$ between
$\mathbf{f}^{-t}_{142}(111)$ and $\mathbf{f}^{-t}_{226}(00)$. Let $x$
be a block of length $m$, $b_0b_1,\ldots,b_{m-1}$. We define
\begin{equation} [T(x)]_i=x_i+x_{i+1} \mod 2
\end{equation}
for $i=0,1,\ldots,m-2$. $T(x)$ is therefore a block of length $m-1$.
Since $T$ is a block evolution operator of rule 60, relationship
similar to (\ref{r142trafo}) must hold, i.e.
\begin{equation} \label{Ttrafo} T \circ \mathbf{f}_{142}
  =\mathbf{f}_{226} \circ T.
\end{equation}
Let us now assume that $b$ is a $t$-step preimage of {\tt 111} under
$\mathbf{f}_{142}$. This means that $\mathbf{f}^t_{142}(b)=111$. Since
$T(111)=00$, we have $T( \mathbf{f}^t_{142}(b))=00$. Using
(\ref{Ttrafo}) we obtain $\mathbf{f}^t_{226} (T (b))=00$. This means
that if $b$ is a $t$-step preimage of {\tt 111} under
$\mathbf{f}_{142}$, then $T(b)$ is a $t$ step preimage of 00 under100
rule 226.
  
Now let us consider a transformation inverse to $T$. In general, $T$
is not invertible, but if restricted to the set of preimages of 111
under $\mathbf{f}_{142}$, it becomes invertible.

For an arbitrary block $y$, there exist two different blocks $x$ such
that $T(x)=y$, and one can show that these two blocks are related by
Boolean conjugacy.  For example, we have $T(111)=00$ and $T(000)=00$.
We have to define $T^{-1}$ such that this ambiguity is removed. This
can be done as follows. Let $a$ be a $t$-step preimage of 00 under
rule 226, $a=a_0a_1\ldots a_{2t+1}$. We define
\begin{align*}
  [T^{-1}(a)]_{2t+2}&=1\\
  [T^{-1}(a)]_{2t+1}&=1+a_{2t+1} \mod 2\\
  [T^{-1}(a)]_{2t}&=1+a_{2t}+a_{2t+1} \mod 2\\
  \cdots\\
  [T^{-1}(a)]_{0}&=1+a_0+a_1+ \ldots+ a_{2t+1} \mod 2,
\end{align*}
or in a general form
\begin{equation} [T^{-1}(a)]_i = \begin{cases} 1 + \displaystyle
    \sum_{j=i}^{2t+1} a_j \mod 2 &
    \text{ if $i=0,1,\ldots,2t+1$},\\
    1 & \text{if $i=2t+2$}
  \end{cases}.
\end{equation} 
One can easily show that the above transformation is indeed an inverse
of $T$, and in addition we guarantee that when $a$ ends with two
zeros, $T^{-1}(a)$ ends with three ones, are required for an $t$-step
preimage of 111 under rule 142.
   
Now, if $a$ is a $t$-step preimage of 00 under rule 226, we have
$\mathbf{f}^t_{226}(a)=00$, hence
$\mathbf{f}^t_{226}(T(T^{-1}(a)))=00$, and by eq. (\ref{Ttrafo}) we
obtain $T(\mathbf{f}^t_{142}(T^{-1}(a)))=00$.  The last equation
implies that $\mathbf{f}^t_{142}(T^{-1}(a))=111$, which means that
$T^{-1}(a)$ is an $t$-step preimage of 111 under rule 142, as
required. $\square$

Proposition~\ref{preimenum} follows form the above result and
Proposition~\ref{fuks97prop}(ii).

\section{Probability of occurrence of 111}
The bijective transformation $T$ constructed in in the proof of
Proposition~\ref{equival} has a property which will be useful in
computing $P_t(111)$. Let us call the block $x=x_0x_1$ a
\emph{matching pair} if $x_0=x_1$, and a \emph{mismatched pair} if $x_0
\neq x_1$.  If $a=T(b)$, then the number of matching pairs in $b$ is
equal to the number of zeros in $a$, while the number of mismatched
pairs in $b$ is equal to the number of ones in $a$. This fact,
together with Proposition~\ref{fuks97prop} immediately leads to the
conclusion that under the rule 142, the number of $t$-step preimages
of 111 with exactly $n_0$ matching pairs and $n_1$ mismatched pairs is
equal to
\begin{equation}
  \frac{n_0-n_1}{n_0+n_1}  \binom{n_0+n_1}{n_1},
\end{equation}
where $n_0+n_1=2t+2$. Probability of occurrence of a matching pair in a
in the initial configuration drawn from the initial distribution is $2
\rho$, and the mismatched pair is $1-2\rho$. Therefore, the
probability of occurrence of a block with prescribed sequence of
matching and mismatched pairs such that it has exactly $n_0$ matching
pairs and $n_1$ mismatched pairs is equal to $(2\rho)^{n_1}
(1-2\rho)^{n_0}$. This implies that the probability that a block of
length $2t+3$, randomly selected from the distribution (\ref{initdist}),
is a $t$-step preimage of 111 with exactly $n_0$ matching pairs is
equal to
\begin{align} \label{probn0} \left(\frac{1}{2}\right)
  \frac{n_0-n_1}{n_0+n_1} \binom{n_0+n_1}{n_1} (2\rho)^{n_1}
  (1-2\rho)^{n_0} =\nonumber \\ \frac{n_0-n_1}{4t+4} \binom{2t+2}{n_1}
  (2\rho)^{n_1} (1-2\rho)^{n_0}. 
\end{align}
The factor $1/2$ in front comes from the fact that there are always
two strings with a given sequence of pairs (related by Boolean
conjugacy), but only one of them is a preimage of 111.
  
The smallest possible number of matching pairs in a $t$-step preimage
of 111 is $t+2$ (recall that the number of matching pairs must exceed
the number of mismatched pairs), while the maximum possible number is
$2t+2$ (all zeros).  Summing (\ref{probn0}) over $n_0$ we obtain
\begin{equation}
  P_t(111)=\sum_{n_0=t+2}^{2t+2}
  \frac{n_0-(2t+2-n_0)}{4t+4}  \binom{2t+2}{2t+2-n_0} (2\rho)^{2t+2-n_0} 
  (1-2\rho)^{n_0}.
\end{equation} 
Introducing a new summation index $j=n_0-(t+1)$ we get
\begin{equation} \label{p111} P_t(111)=\sum_{j=1}^{t+1} \frac{j}{2t+2}
  \binom{2t+2}{t+1-j} (2\rho)^{t+1-j} (1-2\rho)^{t+1+j},
\end{equation}  
and as a result, the probability (\ref{Probdis2}) becomes
\begin{equation}
  P_\mathrm{dis}(\rho,t)=1-2\rho-\sum_{j=1}^{t+1}
  \frac{j}{t+1}  \binom{2t+2}{t+1-j} (2\rho)^{t+1-j} (1-2\rho)^{t+1+j},
\end{equation}  
where $\rho \in [0,1/2]$.

\section{Equilibrium probability}
We will now show how to obtain the equilibrium probability, i.e., $\lim_{t
  \to \infty} P_\mathrm{dis}(\rho,t)$. 
 In order to find the limit $\lim_{t
  \rightarrow \infty} P_t(111)$ we can write eq. (\ref{p111}) in the
form
\begin{equation} \label{binomform}
  P_t(111)=\sum_{j=1}^{t+1}\frac{j}{2t+2} b(t+1-j,2 (t+1),2\rho),
\end{equation}
where
\begin{equation}
  b(k,n,p)=\binom{n}{k} p^k (1-p)^{n-k}
\end{equation}
is the distribution function of the binomial distribution. Using
de~Moivre-Laplace limit theorem, binomial distribution for large $n$
can be approximated by the normal distribution
\begin{equation} \label{demoivre} b(k,n,p)\sim\frac{1}{\sqrt{2 \pi
      np(1-p)}} \exp{\frac{-(k-np)^2}{2np(1-p)}}.
\end{equation}
To simplify notation, let us define $T=t+1$. Now, using
(\ref{demoivre}) to approximate $b(T-j,2 T,2\rho)$ in
(\ref{binomform}), and approximating sum by an integral, we obtain
\begin{equation}
  P_{t}(111)=\int_{1}^{T} \frac{x}{2T}
  \frac{1}{\sqrt{8 \pi  T \rho(1-2\rho)}}
  \exp{\frac{-(T-x-4 T \rho)^2}{8 T \rho(1-2\rho)}} dx.
\end{equation}
Integration yields
\begin{align*}
  P_{t}(111)&=\\ 
  &\sqrt{\frac{\rho (1-2\rho)}{2 \pi T}}
  \left\{\exp\left(\frac{-(1-T+4 \rho T)^2}{8 T
        \rho (1-2\rho)}\right) - \exp\left({\frac{-2 \rho T
        }{2 (1-2\rho)}}\right) \right\} +\\
  &\frac{1}{4}(1-4\rho)  \left\{
    {\rm erf}\left(\frac{4 \rho T}{\sqrt{8 \rho (1-2\rho) T}}\right) -
    {\rm erf}\left(\frac{1- T + 4 \rho T}{\sqrt{8 \rho (1-2\rho)
          T}}\right)
  \right\},
\end{align*}
where ${\rm erf}(x)$ denotes the error function
\begin{equation} {\rm erf(x)}=\frac{2}{\sqrt{\pi}} \int_{0}^{x}
  e^{-t^2} dt.
\end{equation}
The first term in the above equation (involving two exponentials)
tends to~$0$ with~$T \rightarrow \infty$. Moreover, since $\lim_{x
  \rightarrow \infty}{\rm erf}(x)=1$, we obtain
\begin{eqnarray*}
  \lim_{t \rightarrow
    \infty} P_t(111)=\frac{1}{4}(1-4\rho)
  \left\{ 1-
    \lim_{T \rightarrow \infty} {\rm erf}\left(\frac{1- T +
        4 \rho T}{\sqrt{8 \rho (1-2\rho)
          T}}\right)
  \right\}.
\end{eqnarray*}
Now, noting that
\begin{equation}
  \lim_{T \rightarrow \infty}
  {\rm erf}\left(\frac{1- T + 4 \rho T}{\sqrt{8 \rho (1-2\rho)
        T}}\right)=
  \left\{ \begin{array}{ll}
      1,  & \mbox{if $4 \rho \geq 1$}, \\
      -1,    & \mbox{otherwise},
    \end{array}
  \right.
\end{equation}
we obtain
\begin{equation}
  \lim_{t \rightarrow \infty} P_t(111)=
  \left\{ \begin{array}{ll}
      1/2-2\rho  & \mbox{if $\rho<1/4$}, \\
      0    & \mbox{otherwise}.
    \end{array}
  \right.
\end{equation}
The final expression for the equilibrium probability becomes
\begin{equation} \label{finalres}
  \lim_{t \rightarrow \infty} P_\mathrm{dis}(\rho,t)=
  \left\{ \begin{array}{ll}
      2\rho  & \mbox{if $\rho<1/4$}, \\
      1-2\rho    & \mbox{otherwise}.
    \end{array}
  \right.
\end{equation} 
\section{Current}
The equilibrium probability calculated in the the previous
section exhibits a singularity at $\rho=1/4$. This singularity
is of a similar nature as the jamming transition observed
in CA rules 184, 226, and related models. 

Recall that in section 3 we defined the current $J$ (eq. \ref{currentdefinition}).  
The expected value of the current is  $i$-independent, so we can define
 the expected current as
\begin{equation} \label{defexpcurrent} j(\rho,t)=E_\mu
  \big[J(s_{i}(t),s_{i+1}(t),s_{i+2}(t))\big] = E_\mu \left[ s_{i}(t)
    s_{i+1}(t) (s_{i+2}(t)-1) \right].
\end{equation}
The graph of $j(\rho,\infty)$ as a function of $\rho$ is known as
\emph{fundamental diagram}.
Using the notion of block probabilities we can rewrite
(\ref{defexpcurrent}) in an alternative form as
\begin{equation}
  j(\rho,t)=-P_t(110),
\end{equation} 
and using (\ref{pdis1})
\begin{equation}
  j(\rho,t)=-\frac{1}{2} P_\mathrm{dis}(\rho,t).
\end{equation} 
The probability of having a dissenting neighbour, as we can see,
is proportional to the expected current. 

Since the current $J$ represents the flow of blocks
$10$, the expected current must be equal to
\begin{equation}
 j(\rho,t)=\rho v(\rho,t),
\end{equation} 
where $v(\rho,t)$ is the expected velocity of a block $10$
at time $t$. Using (\ref{finalres}) this velocity is given by

\begin{equation} 
  \lim_{t \rightarrow \infty} v(\rho,t)=
  \left\{ \begin{array}{ll}
      -1  & \mbox{if $\rho<1/4$}, \\
      1-\frac{1}{2\rho}    & \mbox{otherwise}.
    \end{array}
  \right.
\end{equation} 
We can see that for densities of blocks $10$ smaller than $1/4$,
the average velocity remains constant and equal to $-1$,
which means that all blocks are moving to the left. 
At $\rho=1/4$ a \emph{jamming transition} occurs, and
when $\rho$ increases beyond $1/4$, more and more blocks
are stopped. This phenomenon is very similar to 
jamming transitions in discrete models of traffic flow,
which have been extensively studied in recent years
(\cite{helbing01} and references therein).

\section{Conclusions}
We investigated dynamics of the cellular automaton rule 142.
It can be transformed into rule 226 by 
a surjective transformation, which turns out to be invertible
if restricted to preimages of $111$. This transformation
allows to compute the probability of having a dissenting
neighbour, which, in turn,  allows to compute  the expected current of blocks $10$.
Rule 142 exhibits jamming transition similar to transitions
occurring in discrete models of traffic flow.

It is worth mentioning that there are other CA rules
conserving the number of blocks $10$ which also exhibit
singularities of fundamental diagrams, for example rules 35 
and 14, as reported in \cite{paper23}. For these rules,
however, there exist no transformation relating them
to other rules with singularities, thus the method
presented in this paper cannot be easily applied.
Nevertheless, the nature of singularities in these
rules appears to be the same, thus some relationship
between them and rules 184/226 may exist. This problem is
currently under investigation and will be reported elsewhere.

\vskip 1cm
\noindent \textbf{Acknowledgements:} The author acknowledges financial
support from NSERC (Natural Sciences and Engineering Research Council
of Canada) in the form of the Discovery Grant.

\providecommand{\href}[2]{#2}\begingroup\raggedright\endgroup

\end{document}